\documentclass[twocolumn,secnumarabic,amssymb,nobibnotes,aps,
prl,
10pt,superscriptaddress
]{revtex4-2}

\usepackage[T1]{fontenc} 
\usepackage[utf8x]{inputenc}
\DeclareUnicodeCharacter{"2061}{}
\usepackage[hidelinks,colorlinks=true,linkcolor=blue,urlcolor=blue,citecolor=blue,]{hyperref}
\usepackage{amsmath} 
\usepackage{cleveref}

\usepackage{graphicx}

\usepackage{float}

\begin{document}

\title[]
  {Observation of Anomalous Josephson Effect in Nonequilibrium Andreev Interferometers}
\author{Daniel Margineda}
\affiliation{NEST Istituto Nanoscienze-CNR and Scuola Normale Superiore, I-56127, Pisa, Italy}
\email{daniel.margineda@nano.cnr.it}
\affiliation{Croton Healthcare. Coral Springs, Florida, 33065, USA}


\author{Jill S. Claydon}
\author{Fatjon Qejvanaj }
\author{Chris Checkley }
\affiliation{Croton Healthcare. Coral Springs, Florida, 33065, USA}

\begin{abstract}

The evidence of anomalous Josephson effect in superconducting-normal-superconducting (SNS) junctions requiring neither Zeeman splitting nor spin-orbit coupling is reported. We demonstrate that a spontaneous phase emerges by modifying the nonequilibrium conditions of the weak link resembling $\varphi_0$-junctions. The voltage-controlled phase shift is detected from magnetoresistance measurements of the proximitized wire connected to the weak link forming a crosslike Andreev interferometer. The interplay of Aharanov-Bohm-like and conventional Josephson currents, in agreement with recent predictions, provides the ingredients for the anomalous current-phase relation. These results represents a breakthrough for engineering superconducting nanocircuits with voltage-tunable quantum properties.

\end{abstract}

\maketitle
The dc Josephson effect establishes that current flows without dissipation across two superconductors interrupted by a weak link in a so-called Josephson junction (JJ)~\cite{jos62}. The supercurrent $I_s$ and the macroscopic phase difference of the superconductors $\delta$ are correlated by the current-phase relation (CPR) which hosts some general properties~\cite{lik79,gol04} like 2$\pi$ periodicity and time-reversal symmetry $I_s (-\delta)=-I_s (\delta)$. The latter determines the $I_s(\delta=0)=0$ condition and a CPR given by a series of sinusoidal functions, normally well described by the first harmonic $I_s(\delta)=I_c \sin{\delta}$ with minimum energy at $\delta=0$ and therefore called 0-JJs. Under certain conditions, the supercurrent can change its polarity leading to a $\pi$ ground state. Discovered in nonequilibrium-controlled metallic JJs~\cite{bas99,sha00}, $\pi$-JJs were intensely investigated thereafter in ferromagnetic~\cite{hei00,rya01}, semiconductor-based JJs~\cite{ke19,dam06} or using superconductors with unconventional paring symmetry~\cite{sch00,smi02}. However, these JJs still present a rigid ground state given by the supercurrent direction which constrains the flowing Josephson current to $\delta \neq n \pi$ states with $n$ any integer. To date, two mechanisms have been explored to achieve JJs with variable ground states. Those with a large and negative second harmonic that gives rise to a degenerate and arbitrary design-determined $\pm \varphi$ phase~\cite{gol07,sic12,gol13}; and junctions with broken inversion symmetry and a non-degenerate and controllable $\varphi_0$ phase bias acting as a Josephson phase battery~\cite{szo16,ass19,str20,may20}. The former is possible in a combination of 0-$\pi$-JJs owing to spatial oscillations of the Cooper pair wave function~\cite{min98,buz03}, ferromagnetic weak links are the most suitable candidates thanks to the ground state dependence on the magnetic layer thickness~\cite{wei06,gur10}. In $\varphi_0$-JJs, the interplay of a strong spin-orbit coupling (SOC) and an exchange splitting field 
~\cite{buz08,ber15,dol15} or noncoplanar magnetic texture~\cite{sil17} induces a finite phase shift
\begin{equation}
I_s=I_c \sin{(\delta -\varphi_0)}=I_J  \sin{\delta} +I_{an} \cos{\delta}
\label{eqn:1}
\end{equation}
with $I_J=I_c \cos{\varphi_0}$ and $I_{an}=-I_c \sin \varphi_0$ the usual and anomalous Josephson current.
		
In diffusive metallic junctions, the amplitude and direction of the Josephson current is given by the occupation of the supercurrent-carrying density of states (SCS) travelling parallel (positive) to the phase gradient, and those carrying the supercurrent in the opposite direction (negative)~\cite{wil98,hei02}. Unlike previous systems, $\pi$-states are achieved by driving the electrons out-of-equilibrium in multiterminal normal-superconducting heterostructures usually called Andreev interferometers~\cite{bas99,sha00,bas02}. An electrostatic potential applied across a mesoscopic conductor coupled to the weak link can provide the nonthermal distribution~\cite{pot97} that allows selective depopulation of the low-energy positive states and hence reversal of the supercurrent. A recent revision of the quasiclassical Green's function formalism suggests that geometric asymmetries in the metal part may induce dissipative currents that gets converted into voltage-dependent Aharonov-Bohm-like supercurrents at the SN interface~\cite{dol18,dol19,dol19S}. The interplay of topology-dependent, even-in-$\delta$ Aharonov-Bohm oscillations and conventional odd-in-$\delta$ Josephson currents might result in nonequilibrium $\varphi_0$-states. However, no such states have been reported so far.

In this letter, we report on the demonstration of nonequilibrium $\varphi_0$-states in metallic junctions forming mesoscopic crosslike Andreev interferometers known as Hybrid Quantum Interference Device (HyQUID)~\cite{hyq,con21}. The anomalous phase is obtained from phase-dependent magnetoresistance of the proximitized normal wire connected to the hybrid junction which is embedded in a superconducting loop with non-negligible screening currents. Three regimes are detected depending on the electrostatic potential. Below a critical voltage, the CPR is governed by the conventional Josephson currents. Anomalous currents dominate the supercurrent at large voltages resulting in $\pi/2$-states. For intermediate values, voltage-tunable $\varphi_0$-states arise.

A single-junction interferometer configuration is effectively used to measure the anomalous phase~\cite{gua20} using three simple arguments: 1) proximity effect on the connecting wire 2) non-negligible screening currents in the superconducting loop and 3) nonequilibrium conditions to manipulate the density of states. The metal conductance of the coupled connector oscillates as a function of the phase difference due to phase transfer from the superconducting condensate to normal electrons via Andreev reflections at the SN interfaces~\cite{pet95} with a minimum resistance value at $\delta=2n \pi$ and equals the normal-state resistance $R_N^{\,0}$ at $\delta=(2n +1) \pi$. Thus, the magnetoresistance can be approximated as~\cite{pet98}
\begin{equation}
R_N=R_N^{\,0}-\frac{\Delta R}{2} [ 1+\cos\delta]
\label{eqn:2}
\end{equation}
For large screening currents in the superconducting loop, the phase difference depends on the supercurrent and becomes a multivalued function of the external flux due to the supercurrent-induced flux. For a certain flux values, the system overcomes a phase-jump transition to a more favorable states with sharp changes on the magnetoresistance called the bifurcation mode~\cite{lik86,con21}. In the presence of an anomalous phase, the magnetoresistance depends on the sign of the external flux breaking the even-in-flux symmetry. Thus, the phase shift can be determined from the hysteretic behavior of the magnetoresistance, without the need to incorporate a second junction in a DC-SQUID configuration.
	
\begin{figure*}[ht]
\includegraphics[scale=0.17 ]{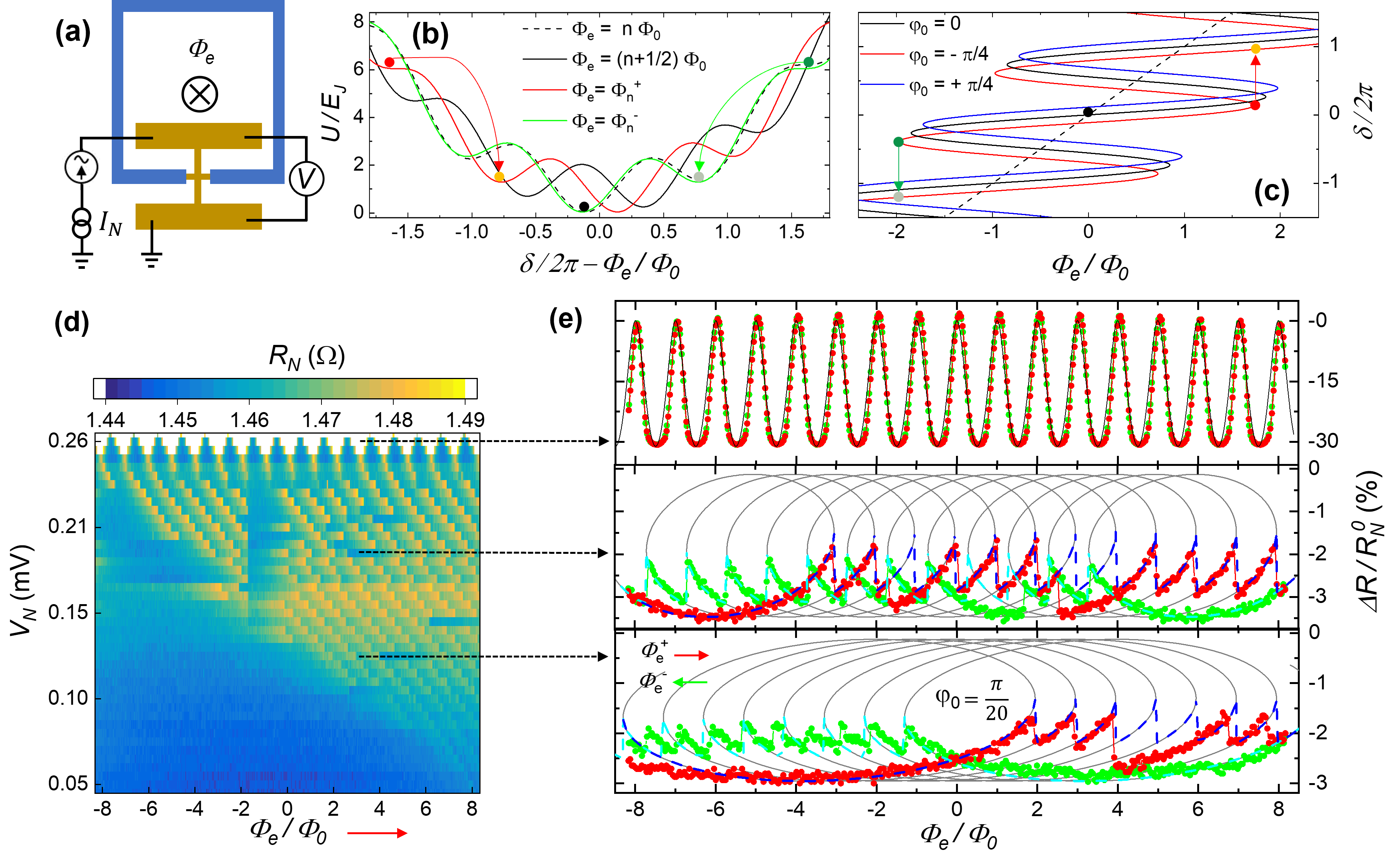}
\caption{(a) HyQUID schematic with the 4-wire electrical configuration for measuring the interferometer (dark yellow) differential resistance $R_N$. (b) Flux-tilted washboard potential of the system for $\varphi_0=-\pi/4$ and $\beta$= 10. Positive (negative) external flux shifts to the left (right) the $\Phi_0$-periodic potential and the particle escapes to a lower energy state at $\Phi_n^+ (\Phi_n^-)$, red (green) balls. (c) $\delta(\Phi_e)$ for  $\varphi_0$ = 0, -$\pi$/4, $\pi$/4. with phase jumps at $d\Phi_e/d\delta=0$. (d) Colormap plot of a HyQUID differential resistance for positively sweep flux and bias voltages $V< V_N^{c1}$. (e) $R_N (\Phi_e)$ traces for selected voltages with positive (red) and negative (green) flux polarity. $ R_N^+ (\Phi_n^+)>R_N^- (\Phi_n^-)$ for all the bias voltages. The gray curves are the best fit to the model. Dashed lines account for phase jump transitions.}
\label{samples} 
	\end{figure*}

We fabricated Nb/Ag/Nb HyQUIDs depicted in Fig.~\ref{samples}a which consist of a 300 nm-thick superconducting loop interrupted by 50 nm-thick non-magnetic metal connected to reservoirs by normal wires with total length $L_N\simeq 4$ $\mu$m in a crosslike geometry. The devices presented here are two-dimensional structures in the diffusive and long junction regime, since the superconducting $\xi_0\sim 40$ nm for Nb, and the normal $\xi_N=\sqrt{\hbar D/2\pi k_B T}\simeq 83 $ nm coherence lengths are shorter than the weak link dimensions $L_x=L_y=$ 500 nm. Large reservoirs and diffusion time $\tau_D=L_N^2/D \simeq$ 1.1 ns shorter than the characteristic interaction time $\tau_{\phi} =L_{\phi}^2/D \simeq$ 1.6 ns ensure nonequilibrium conditions. We performed measurements at T = 2.7 K above the temperature at which long-range correlations decay $E_{th}/k_B \sim 0.6$ K in the so called re-entrance effect ~\cite{naz96}. A Thouless energy $E_{th}=\hbar D/L_x^2=0.04$ meV and relevant parameters are calculated for a diffusion coefficient $D=150$ $cm^2/s$ and phase-breaking length $L_{\phi}=$ 4.9 $\mu$m estimated in the SupplementaI Material (SM)~\cite{SI} (see, also, references~\cite{vol96,usa70,bel99,pet98_2,win86,hen99,kof07,cas99} therein). The chosen temperature is also justified by the exponential decay of the Josephson currents with T~\cite{dub01} faster than the power-law dependence of the Aharanov-Bohm currents $\propto 1/T$~\cite{cou96,gol97}.
 The phase across the junction is controlled by the out-of-plane magnetic flux $\Phi_e$ piercing the superconducting loop. For a screening parameter $\beta=2\pi LI_c/\Phi_0 >1$, the contribution to the net flux due to the supercurrent-induced flux results in a non-linear phase difference: 
\begin{equation}
\delta=2\pi\frac{\Phi_e}{\Phi_0}-\beta i=\phi_e-\beta i
\label{eqn:3}
\end{equation}
with $i=I_s/I_c$ the normalized supercurrent and $\phi_e \equiv 2\pi \Phi_e/ \Phi_0$ the ``external phase''. The potential energy for such junctions is given by the Josephson energy and the magnetic energy stored in the loop inductance~\cite{lik86}
\begin{equation}
U=E_J \left[1-cos\left(\delta -\varphi_0\right)+\frac{\left(\delta -\phi_e \right)^2}{2\beta}\right]
\label{eqn:4}
\end{equation}
with $E_J=\Phi_0 I_c/2\pi$. If $\beta<1$, $\delta \simeq \phi_e$ and the potential energy has a parabolic shape with a single minimum. For large screening currents $\beta>1$, the system enters the bifurcation mode. The oscillating component gives rise to several minima and maxima according to the positive and negative slopes of $\delta(\phi_e)$ as shown in Fig.~\ref{samples}b,c. $U(\delta)$ presents inflection points for
\begin{equation}
\phi_n^{\pm}=\varphi_0 \pm \left[ \phi_t (\beta) + 2\pi n\right]
\label{eqn:5}
\end{equation}

at which $\delta(\phi_e)$ slopes change sign. $\phi_t=\sin^{-1}⁡[1/\beta]+\sqrt{\beta^2-1}+\pi/2$ (see calculations in the SM~\cite{SI}) varies linearly for large $\beta$. For convenience, we use $\Phi \equiv \phi/2\pi$ notation. Under the classical analogy of a particle moving in a ``washboard potential'', the particle is trapped in a potential minimum U=0. When an external flux is applied, the parabolic part is shifted along the oscillating component and the particle energy increases captured in the potential well. The well barriers decrease with the flux until vanishing at $\Phi_e=\Phi_{n=0}^\pm$ entering the bifurcation mode. The induced supercurrent $I_s=\Phi_n/L$ matches the critical current, a flux quantum enters the loop reducing $|\phi_e-\delta|$, and the particle escapes to a lower energy state in a ``phase jump''. In $\varphi_0$-JJs, the zero-flux minimum shifts to $\delta=\varphi_0$ and the anomalous phase breaks the phase difference symmetry $\delta(-\Phi_e) \neq -\delta(\Phi_e)$ with $\phi_n^+= -\phi_n^-+2\varphi_0$. Thus, $R_N(\Phi_e)$ becomes tilted and direction-dependent.

The differential magnetoresistance $R_N$ of the coupled connector is measured as a function of the voltage at the center of the wire given by $V_N=I_NR_N(0)/2$. $I_N$ is the bias current and $R_N(0)$ the differential resistance at zero bias current. All devices exhibit similar resistance $R_N(0)=1.8(2)\Omega$. The flux was swept back(-) and forth(+) to detect the presence of the anomalous phase. $R_N^+(\Phi_e,V_N)$ colormap and selected traces are shown in Fig.~\ref{samples}d,e for $V_N$ < $V_N^{c1}$ = 0.26 mV. Gray curves are the best fit to the model given by Eq.~\ref{eqn:1},~\ref{eqn:2} and~\ref{eqn:3} and phase jumps (dashed trajectories) obtained from $\Phi_n^\pm$. $\Delta R, \beta, \varphi_0$ are fitting parameters. A quantitative study of $\Delta R(V_N)$ is given in the SM~\cite{SI}. The technique account for any trapped flux by referencing the phase jumps to the resistance minimum $\delta=0$. For large $\beta$, the resulting state of the phase jumps are not the one with the lowest energy, several flux quanta may enter the loop and the system falls to a lower energy state or materialize at $\Phi^*<\Phi_n$. These phenomena are observed for almost any voltage regardless of the flux polarity. Notwithstanding the above, the model adequately tracks these anomalies since the system remains trapped in the next available potential well as the gray lines shown. $\Phi_{n=0}$ decreases monotonically until the critical current is suppressed at $eV_N^{c1}\simeq 7 E_{th}$ where the system becomes fully periodic as expected in junctions dominated by Josephson currents. However, the resistance at phase jumps is slightly lower when the flux is negatively swept $R_N^+(\Phi_n^+)> R_N^-(\Phi_n^-)$, explained by a small and quasi-constant anomalous current $I_{an}<0$ corresponding to $\varphi_0= +\pi/20$. A precision $\epsilon (\varphi_0) \leq \pi/40$ is achieved for a flux sampling $\Delta \Phi_e \simeq 20$ m$\Phi_0$ and resistance precision $\simeq$ 1 m$\Omega$.

\begin{figure*}[ht]
\centering
\includegraphics[scale=0.17]{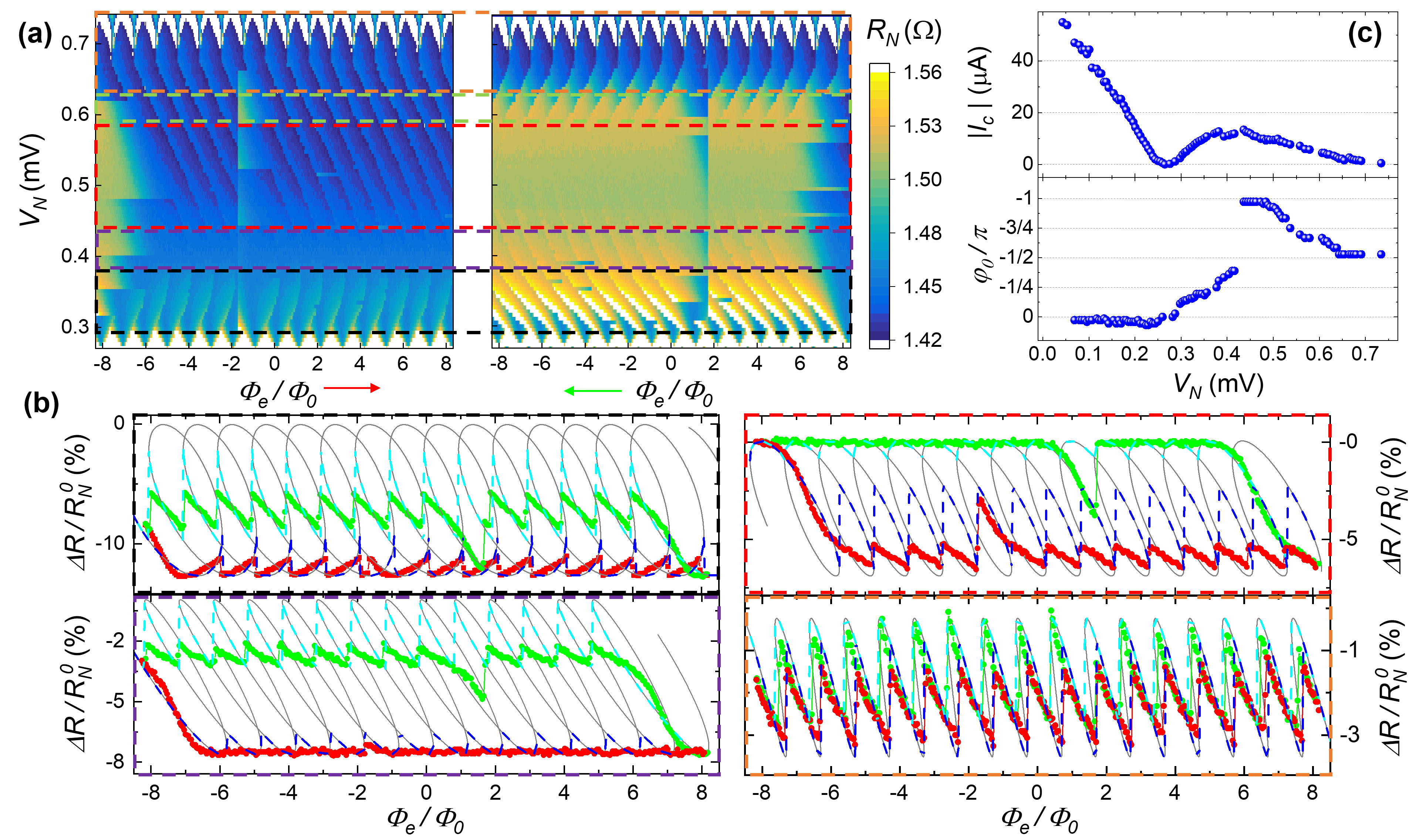}
\caption{(a) Colormap plot of the HyQUID differential resistance for positive (left) and negative (right) sweeping flux polarity above $V_N^{c1}$. (b) Characteristic $R_N^+$ (red) and $R_N^-$ (green) traces for each zone (dashed squares) at $V_N$= 0.33, 0.4, 0.56 and 0.67 mV. $\varphi_0(V_N)$ tilts the magnetoresistance that becomes hysteretic. (c) Effective critical current $\vert I_c \vert$ and $\varphi_0$-states as a function of the bias voltage. The Josephson current $I_J$ reverses at $V_N^* \simeq$ 0.44 mV with a jump discontinuity according to the model. The system eventually tends to a state dominated by the anomalous Josephson current with $\vert I_c \vert \rightarrow 0$ and $\varphi_0=-\pi/2$.}
\label{FB2}
	\end{figure*}
	
Above $V_N^{c1}$, the magnetoresistance becomes strongly hysteretic as shown in Fig.~\ref{FB2}a,b, owing to non-negligible screening currents and a voltage-dependent anomalous phase $\varphi_0(V_N)$. $R_N^{\pm}(\Phi_e)$ shows a sawtooth pattern with jumps from high to low values for both flux polarities. As the bias voltage increases, the resistance jumps in $R_N^-$ do not drop to the lowest resistance state and the curves become hysteretic. The height of the resistance jumps progressively decreases in $R_N^+$ and the gap between the curves increases, well described by a monotonic increase of both the critical current and a negative phase shift $0<|\varphi_0 | <\pi/2$. The sign change in $\varphi_0$ means that only $I_{an}$ is reversed. At $V_N$ = 0.38 mV, $R_N^+$ jumps disappear at $\varphi_0=-\pi/4$ and $\beta \simeq 11$. Beyond this point (purple zone), $R_N^+$ reaches the minimum value and remains constant with some traces of small jumps from low to high resistance states, meanwhile $R_N^-$ jumps are progressively rounded. The model predicts small phase jumps in $R_N^+$ for larger $|\varphi_0 | $ with a flip in the direction of the resistance jumps at $\varphi_0 \simeq -\pi/3$ and large screening currents. Above $V_N^*\simeq$ 0.44 mV, $R_N^-$ peaks disappear, the resistance increases smoothly and remains constant, signature of  the $I_J$ inversion, $\pi >|\varphi_0 | > \pi/2$. Phase jumps in $R_N^-$ are expected to disappear for $\varphi_0=-l \:\pi/6$ and $\beta > 5$ with $l \in 4-5 $ as it is shown for $V_N$ = 0.56 mV. The onset of $R_N^-$ resistance jumps in the green zone with rounded maxima is explained by the decrease of $|\varphi_0|$ and the screening currents that pull the $R_N$ branches apart for $\beta$ < 5 and $\varphi_0\rightarrow - 2\pi/3 $. $R_N^-$ recovers the sawtooth shape with a similar profile than $R_N^+$ for $V_N$ > 0.63 mV, consistent with a CPR controlled by $I_{an}$, which means $\varphi_0 \simeq - \pi/2$. The gap decreases and the system becomes periodic and fully symmetric $R_N^-=R_N^+$ for $V_N$ > 0.68 mV. For $|\varphi_0|=\pi$, $R_N$ is not tilted and the resistance branches cross near their maximum value at a given flux $\Phi^*(\beta )$. For $V_N \in V_N^*-$ 0.5 mV, we assumed that the escape rate from the potential well is non-negligible at $\Phi^* <\Phi_n$ and the model can fit the data considering the initial curvature and the $R_N^+$ jumps. The fitting parameters are plotted in Fig.~\ref{FB2}c with a jump-like transition to quasi-$\pi$ states at $V_N^*$ Thermallly activated phase jumps (TAPJ) around $V_N^*$ are discussed in the SM.~\cite{SI}.

 \begin{figure}[ht]
\includegraphics[scale=0.11]{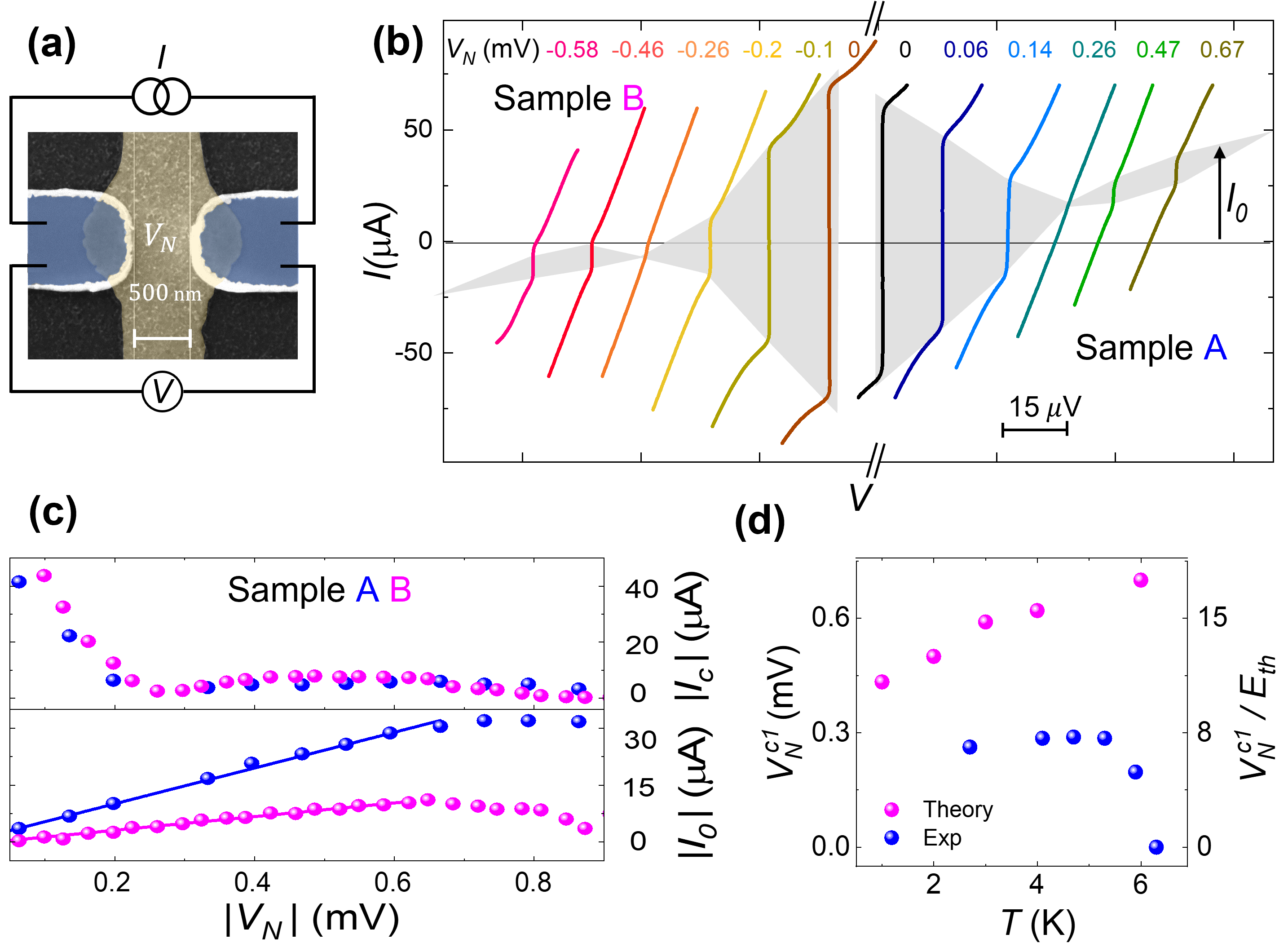}
\caption{(a) False-colored micrograph of the junction (superconducting electrodes in blue) with the electrical configuration to measure the differential resistance. (b) Current-voltage characteristics for selected voltages at the center of the metal $V_N>0$ (sample A) and  $V_N<0$ (sample B). Superconducting region $I_s$ is marked in gray shifted by  a dissipative current $I_0$. The curves are horizontally offset for clarity. (c) Characteristic parameters of the anomalous junctions. The sample-dependent $I_0$ scales linearly for $V_N< 0.7$ mV. (d) Comparison of experimental and theoretical $V_N^{c1}$ values}
\label{CB} 
	\end{figure}
	
Differential resistance of the junction in HyQUIDs without the superconducting loop allows direct probing of the critical current and the nonequilibrium conditions. SN interfaces exhibit high transparency since normal state resistance $R_{sns}$ = 0.24(1) $\Omega$ is similar to the one estimated from resistivity $R_n$ = 0.22(1) $\Omega$.  The I-V characteristics, obtained by integration, for selected $V_N$ values applied as before are shown in Figure~\ref{CB}b. $I_c$ decreases mimicking the results obtained in flux-biased HyQUIDs with a fully resistive state at a similar critical value $V_N^{c1}$. The persistent current emerges for higher voltages until it vanishes definitively at $eV_N^{c2}\simeq$ 0.9 meV $\simeq$ 22 $E_{th}$ confirming nonequilibrium conditions $\tau_D<\tau_{\phi}$. Otherwise, the critical current monotonically decay without reversing in a hot electron regime~\cite{mor98}.
 A dissipative and voltage-dependent current $I_0$ displaces the superconducting region $I_s$ (gray area) as predicted in asymmetric interferometers~\cite{dol19S,dol19}. $I_0$ scales linearly with the applied voltage up to $V_N\simeq0.7$ mV, beyond which thermal effects undermine phase-coherence correlations as shown in long junctions~\cite{SI}. Its intrinsic nature is confirmed by reversing the polarity of the bias voltage in sample B. In case any leakage current flows from the control wire to the junction ground, the dissipative current would remain positive and the critical voltage should differ from the one obtained in magnetoresistance measurements, contrary to the observations.

The temperature dependence of the critical voltage is compared with the existing theory for symmetric interferometers~\cite{wil98,hei02}. The electrostatic potential depopulates states with $\epsilon< eV_N$. Temperature undermines the selectivity of the depopulation of positive states increasing $V_N^{c1}$ until the supercurrent is no longer reversed. In our HyQUIDs, the critical voltage remains constant with temperature up to T $\geq$ 6 K. The expected values are plotted together with the experimental values in Fig.~\ref{CB}d. Details of the calculations in the SM~\cite{SI}. The fact that the theoretical $V_N^{c1} \simeq V_N^*$ suggests that the anomalous current modifies the SCS and reduces the critical voltage.

Spin-orbit interactions and/or a non-coplanar spin structure can be ruled out as the driving force since we used non-magnetic metals and the characteristic Kondo upturn caused by magnetic impurities in $R_N(0,T)$ was not observed. Unconventional pairing symmetries are not expected in Nb-based devices and the evolution of the differential resistance cannot be fitted if higher harmonics are considered. Nevertheless, our results are fairly explained by the competition of Josephson $I_J$ and Aharanov-Bohm-like $I_{AB}$ currents. The dissipative currents becomes phase-dependent at the SN interfaces, adding a new term $I_{AB}$ to the CPR with an even-in-$\delta$ parity. While $I_J(V_N)$ decays exponentially, $I_{AB}(V_N)$ increases to saturation at a certain voltage~\cite{dol19S}, explaining the evolution from quasi-zero- to $\pi/2-$states and the interesting $\varphi(V_N)-$ states at intermediate voltages. 
We can quantify the asymmetries of our HyQUIDs based on the junction geometry. Two scenarios with anomalous Josephson effect 
were investigated: one with nonsymmetric cross branches~\cite{dol19} and one with a nonpointlike cross intersection~\cite{dol18,dol19S}. The former geometry requires different lengths from the center of the cross to the superconducting electrodes, but also to the metallic reservoirs. Although $L_y=L_x$ can account for the first condition $L_y/L_N\simeq$ 1/8 is rather small. The latter geometry is less restrictive and the width and irregular interfaces shown in Fig.~\ref{CB}a meet the requirement. Measurements on narrower HyQUIDs, $L_y\simeq$ 150 nm but similar SN interfaces and transparency still exhibit $\pi/2-$states, see SM~\cite{SI}. Thus, we can conclude that $L_y/L_x\simeq 0.3$ provides sufficient asymmetry to promote the anomalous currents.

In summary, our work reveals the first evidence of anomalous Josephson effect in metallic and nonmagnetic SNS junctions configured as Andreev interferometers in which time-reversal and inversion-symmetries are broken due to the metal topology. Our model provides a route to unambiguously determine the anomalous phase if large screening currents are present. $\varphi_0$-states can be tuned by modifying the metal electrostatic potential and probed in flux-biased junctions. HyQUIDs do not require side gates to control the electron density in semiconductor-based $\varphi_0$-junctions, complex 2DEG systems or scarce unconventional superconductors.

\begin{acknowledgments}
We gratefully acknowledge Prof. V. Petrashov and Prof. G. Green for for fruitful discussions. This work was supported by York Instruments LTD.

\end{acknowledgments}



\bibliography{bibliography_varphi}

\end{document}